\documentclass[twocolumn,showpacs]{revtex4}
\usepackage{graphicx}
\usepackage{amsmath}
\usepackage{amsfonts}
\usepackage{amssymb}
\usepackage{color}
\usepackage{amscd}
\usepackage{bm}

\begin{document}
\title{Some thoughts on constructing a microscopic theory with holographic degrees of freedom}
\author{Yong Xiao}
\email{xiaoyong@hbu.edu.cn}
\address{College of Physical Science and Technology, Hebei University, Baoding 071002, China}

\begin{abstract}
Holographic principle states that the maximum entropy of a system is
its boundary area in Planck units. However, such a holographic
entropy cannot be realized by the conventional quantum field theory.
We need a new microscopic theory which naturally possesses all the
holographic degrees of freedom. In this paper, we provide some
preliminary thoughts on how to construct a theory with holographic
degrees of freedom. It may shed light on the understanding of
quantum properties of gravity and the early stage of the universe.
\end{abstract}
 \maketitle

\section{Introduction}
Holographic principle states that the maximum entropy contained in a
system is its boundary area in Planck units
\cite{hooft,susskind,bousso}. However, the conventional quantum
field theory (QFT) cannot provide enough degrees of freedom (DoFs)
to account for the holographic entropy. As analyzed in \cite{cohen},
the energy and entropy of an ordinary QFT system can be expressed as
$E=L^3 \Lambda^4$ and $S=L^3 \Lambda^3$ where $L$ is the size of
system and $\Lambda$ is the effective ultraviolet cutoff of the
system, which can also be easily obtained by dimensional analysis.
Imposing a requirement that the maximum energy of the system does
not exceed the energy of a black hole of the same size, i.e., $E=L^3
\Lambda^4 \leq L/G$, it immediately leads to $\Lambda \leq
(\sqrt{G}L)^{-1/2}$ which is called the ultraviolet-infrared (UV-IR)
relation in \cite{cohen}. Substituting it into the entropy formula,
one finds that the maximally realizable entropy of a ordinary QFT
system is $S\leq (A/G)^{3/4}$. In a word, though QFT is usually
viewed as theory with infinite DoFs (attached with an unlimited
value of $\Lambda$), after imposing the energy limitation of general
relativity, the maximum realizable entropy becomes $(A/G)^{3/4}$.
This $(A/G)^{3/4}$ entropy bound for conventional QFT has been
obtained and verified in various contexts
\cite{hooft,cohen,hsuEB,us1,barrow,hsuMN}. Throughout the paper, we
reserve the gravitational constant $G$ to highlight the influence of
gravity, while the other fundamental constants $\hbar$, $c$, $K_B$
and those unimportant numerical factors are omitted in most
expressions.

Cohen et al. also introduced another UV-IR relationship $\Lambda
\sim (GL)^{-1/3}$ in \cite{cohen} with the purpose of saturating the
holographic entropy bound, i.e., making $L^3\Lambda^3 \sim A/G$. But
they soon excluded the UV-IR relation because $E=L^3 \Lambda^4\gg
L/G$ in this case. Interestingly, such a kind of $(GL)^{-1/3}$
behavior or its variants can always be found in the literature
\cite{jack1,jack2,Mazia,us2}. A recent example is that the specific
volume of the constitutes of black boles is identified to be $v=G
r_h$ by comparing with the properties and equations of van der Waals
fluids \cite{wei}. This can be easily translated to a length size
$(Gr_h)^{1/3}$ or $(Gr_h)^{-1/3}$ in momentum space. Thus we should
accept $\Lambda = (GL)^{-1/3}$ as a useful UV-IR relation, which
should be applied to a holographic system rather than a conventional
QFT system. Meanwhile, the energy formulae $E=L^3 \Lambda^4$ cannot
be applicable for the case with $\Lambda=(GL)^{-1/3}$. To make it
clear, let us examine the gravitational correction to the energy of
an ordinary QFT system. The self-gravitational potential energy can
be estimated as $G\frac{(L^3 \Lambda^4)(L^3 \Lambda^4)}{L}$ and it
is negligible compared to the energy $E=L^3 \Lambda^4$ for $\Lambda
\ll (\sqrt{G}L)^{-1/2}$. But with $\Lambda> (\sqrt{G}L)^{-1/2}$, the
gravitational correction is too large and make the ordinary QFT
description invalid. So we need a new theory from
$(\sqrt{G}L)^{-1/2}$ to $(GL)^{-1/3}$. Many new DoFs should emerge
in this range and finally overcome the entropy gap from
$(A/G)^{3/4}$ to $A/G$. The situation is visualized as follows.
\begin{align*}
 \begin{CD}
 \Lambda:  0 @>\text{Ordinary QFT}>S\leqslant (A/G)^{3/4}> (\sqrt{G}L)^{-1/2}
 @>\text{New theory?}>S\leqslant A/G > (GL)^{-1/3}.
 \end{CD}
\end{align*}
In this paper, we aim to provide some thoughts on understanding the
physics in the range from $(\sqrt{G}L)^{-1/2}$ to $(GL)^{-1/3}$,
which has never be carefully studied before as far as we know. It is
natural to expect the behavior in this range should be closely
related to the quantum properties of gravity. But due to our lack of
the knowledge of a complete theory of quantum gravity, we mainly
rely on the holographic principle to guide us. We concentrate on
answering such a question: how to construct a microscopic theory
with holographic DoFs and how the theory is distinct from our
familiar QFT. We shall show that such a holographic theory can be
successfully constructed. And from this theory it is easy to derive
the thermodynamical behaviors $E\sim L/G$, $S\sim A/G$ and $T\sim
L^{-1}$ that are typical for black holes.

\section{Lessons from the Debye theory for solids}
Holographic principle imposes a maximum DoFs for any
quantum-gravitational system. And in solid physics, Debye theory
\cite{huang} also has a limitation to the maximum DoFs of the
corresponding system. So here is a suitable place for us to learn
some lessons from the Debye theory.

The Debye model considers a solid as $N$ non-interacting quantum
harmonic oscillators. So the total DoFs for the system is $3N$. To
insure this limitation, the concept of Debye frequency $w_D$ is
introduced and its value can be calculated from the requirement
\begin{align}
L^3 \int_0^{w_D } {w^2 dw = 3N}.\label{wd}
\end{align}
As always, to avoid unnecessary complications, we omit an analysis
of the differences between the longitudinal and transverse DoFs in
the solid. From Eq.\eqref{wd} we get $ w_D \sim N^{1/3} /L$ .
Knowing the Debye frequency, the energy of the system is expressed
as
\begin{align}
 E = L^3 \int_0^{w_D }
{\frac{w}{{e^{w/T} - 1}}w^2 dw} .
\end{align}
For a temperature far below $w_D$, the upper limit of integration
can be approximately extended to $\infty$. Thus the energy is
calculated as
 \begin{align}
 E \approx
L^3 \int_0^\infty  {\frac{w}{{e^{w/T}  - 1}}w^2 dw}  \sim L^3 T^4 .
\end{align}
For a temperature with $T \gg w_D$, there is $e^{w/T} - 1 \approx
w/T$. Then the energy is expressed as
 \begin{align}
 E \approx L^3 \int_0^{w_D } {Tw^2 dw} \sim NT.
\end{align}
The physical picture of Debye model is clear. At low energy scale,
$T\ll w_D$, the system is described by phonon gas which exhibits the
thermodynamical behaviors
 \begin{align}
E=L^3 T^4,\ \ \ \ \  \ \ S=L^3T^3\ll 3N.
\end{align}
It gives an effective description of the situation where the DoFs of
the system are far from being totally excited. In contrast, at the
energy scale larger than $w_D$, the behaviors of the system become
 \begin{align}
E\sim NT,\ \ \ \ \ \ \   S=3N.
\end{align}
In this situation, all the DoFs are excited and the thermodynamics
of phonon gas is no longer a reasonable description of the system.

Now we have learned important lessons from the Debye theory: a
system exhibits very different behaviors at different energy scale.
Back to the problem we concerned, the ordinary QFT and its
thermodynamical behaviors $E=L^3 \Lambda^4$ and $S=L^3 \Lambda^3$
are very similar to the low energy scale behaviors of the Debye
model. The common characteristic is that they are only applicable to
the cases where the DoFs are not fully excited. Then, after some
scale, the physical behavior will be dramatically changed. The
maximum DoFs that can be excited is $A/G$ for a
quantum-gravitational system and $3N$ for a Debye solid, save that
the related physical mechanisms are different in the two kinds of
systems.

\section{A theory with holographic DoFs}
\subsection{Thermodynamical analysis}
A system can exhibit very different behaviors in different energy
scale. We therefor hope to conceive a flexible thermodynamical
formulae applicable to various situations. We suggest that the
thermodynamics of a system can be generally put into the form
\begin{align}
S = L^3 \Lambda ^3, \label{gs}\\
 E = L^3 \Lambda ^3 T.\label{ge}
\end{align}
Here the crucial setting is that we treat $\Lambda$ and $T$
differently, after all they are respectively attached to momentum
and energy. The parameter $\Lambda$ is understood as the effective
moment cutoff. By intuition every Dofs is located in a size of
$\Lambda^{-1}$, so the number of independent elements is $ L^3
/\Lambda ^{-3}$ and it is consistent with the entropy formula. In
contrast, the temperature $T$ is understood as the average energy
distributed to every DoF. We did not introduce the mass parameter
$m$, because a system consisting of relativistic massless particles
always has more entropy and thus is more appropriate for the
analysis of entropy bounds.

Now look at the ordinary QFT case with $E=L^3 \Lambda^4$ and $S=L^3
\Lambda^3$. Compared with Eqs.\eqref{gs} and \eqref{ge}, we find
$T\sim \Lambda$. It reflects the fact that in conventional QFT the
energy and momentum are treated on the same foot. Concretely
speaking, for massless particles of the QFT we always have the
energy-moment relation
\begin{align}
\varepsilon=cp,
\end{align}
where $p=|\vec{p}|$.

Then we want to describe a system with all the holographic DoFs
being exited. To fit with $S=A/G$ and $E=L/G$, we must have
$T=L^{-1}$ and $\Lambda=(GL)^{-1/3}$ in Eqs.\eqref{gs} and
\eqref{ge}. Considering that $T$ and $\Lambda$ are respectively
related to energy and momentum, there should be
\begin{align}
\varepsilon=Gp^3,
\end{align}
rather than $\varepsilon=cp$ as in the ordinary QFT situation. We
call this new type of particles as ``holographic particles". The
relation looks very weird, but it is comprehensible because we
should expect the gravitational constant $G$ plays a central role in
the holographic case. Whatever, it is worthy to mention that we
don't expect a racial change of the energy-moment relation. The
parameter $\varepsilon$ should always be understood as the energy
distributed to a DoF. We will come back to its explanation later in
the paper.

\subsection{The validity of $\varepsilon=Gp^3$ \uppercase\expandafter{\romannumeral01}}
We have found a relation $\varepsilon=Gp^3$ for holographic
particles. Surely we are eager to find something unfamiliar to avoid
going round in the circle of ordinary QFT and finally to capture the
holographic DoFs. But we still need some confidence of the relation
before making further explanation. Next we shall illustrate the
validity of the relation $\varepsilon=Gp^3$. Using
$\varepsilon=Gp^3$ as our starting point, we show that the
holographic thermodynamics $E\sim L/G$, $S\sim A/G$ and $T\sim
L^{-1}$ can be derived consistently.

For photon fields confined in a box of size $L$, the ensemble
approach will lead us to the thermodynamical behaviors
\begin{align}
 E=\frac{\pi^2}{15}L^3 T^4,\ \ \  \
S=\frac{4 \pi^2}{45} L^3 T^3,
\end{align}
which are the standard QFT behaviors.

Now keeping $\varepsilon=Gp^3$ in mind, we have a new microscopic
theory for holographic particles. All the analysis for photon system
can be translated to the new system, expect that the relation
$\varepsilon=cp$ has to be replaced by $\varepsilon=Gp^3$. The
logarithm of the partition function now is
 \begin{align}
 \begin{split}
\ln \Xi &= -\sum\limits_{i}\ln(1-e^{-\beta \varepsilon})
\\&=-\frac{gL^3 }{2\pi ^2 } \int_0^\infty  \ln \left( 1 - e^{ - \beta Gp^3 }  \right) p^2 dp =   \frac{g}{12}\frac{L^3
}{G\beta},
\end{split}
\end{align}
where $\beta\equiv 1/T$ and $g$ represents other DoFs such as
polarizations. Then we get the expressions for the energy and
entropy of the system as
\begin{align}
E = - \frac{\partial }{\partial \beta}  \ln \Xi = \frac{g}{12}
\frac{L^3 T^2}{G},
\end{align}
and
\begin{align}
S =  \ln \Xi  + E/T = \frac{g}{6} \frac{{L^3 T}}{G}. \label{entropy}
 \end{align}
Substituting into $T=G\Lambda^3$, these formulae can also be
reexpressed as
\begin{align}
E \sim GL^3 \Lambda ^6, \ \ \  \  S\sim L^3 \Lambda ^3.
\end{align}
Obviously the system has distinct thermodynamical behaviors from
those of conventional QFT. For example, the system has lower
temperature and larger entropy density compared to a conventional
QFT system with the same energy. When we require the system has an
energy $E\sim L/G$, it immediately follows that $T\sim L^{-1}$ and
$S\sim A/G$. In addition, the Komar mass corresponds to
$(\rho+3p)V$, so we get $M=4E=\frac{g}{3} \frac{L^3 T^2}{G}$ (the
relation $\rho=p$ is derived right away). Comparing with
Eq.\eqref{entropy}, we observe a relation $M=2TS$, which is the same
as the relation for a Schwarzschild black hole.

Thus, using $\varepsilon=Gp^3$ as the starting point, the typical
holographic thermodynamics $E \sim L/G$, $T \sim L^{-1}$ and $S \sim
A/G$ can be obtained.  Note that though the temperature $T$ is of
the order of $L^{-1}$, the corresponding momentum $\Lambda$ is of
the order of $(GL)^{-1/3}$, which was inconspicuous in previous
knowledge of black hole thermodynamics.

\subsection{The validity of $\varepsilon=Gp^3$ \uppercase\expandafter{\romannumeral02}}

We can also calculate the pressure of the system as
\begin{align} p
= - \frac{\partial F}{\partial V} = T\frac{\partial \ln \Xi
}{\partial V} = \frac{g}{12}\frac{T^2 }{G}. \end{align} Comparing
with $\rho=U/V= \frac{g}{12} \frac{{T^2 }}{G}$, we get $\rho =p$
\cite{note}. Fundamentally this behavior comes from the fact that
$\varepsilon_i=Gp_i^3= \frac{1}{V} \sqrt {n_x ^2  + n_y ^2  + n_z ^2
}\sim \frac{1}{V}$, where the momenta are quantized as
$\vec{p}=\frac{1}{L} (n_x,n_y,n_z)$. Thus there is naturally
\begin{align}
P =  - \sum\limits_i {a_i } \frac{{\partial \varepsilon _i
}}{{\partial V}} =\sum\limits_i {a_i \frac{\varepsilon _i}{V} } =
\frac{U}{V}= \rho.
\end{align}
Interestingly, Fischler and Susskind \cite{fs} applied the
holographic principle to cosmology and found that in the flat FRW
universe case the holographic entropy bound can only be saturated
with the equation of state $p=\rho$. Banks and Fischler
\cite{bf1,bf2,bf3,bf4} contributed many efforts on the holographic
cosmology with $p=\rho$ which can dismiss the Big Band singularity,
and they further proposed a holographic eternal inflation model by
noting that a black hole with dS interior can be embedded in a
$p=\rho$ background. Moreover, the $p=\rho$ fluid has also been used
to construct stable dense stars as the endpoints of gravitational
collapse, which have no event horizons and no singularities
\cite{mm}. We think the natural derivation of $\rho=p$ in our model
as an evidence of the validity of $\varepsilon=Gp^3$.

\section{The field-theoretical viewpoint}
Now we have a microscopic theory from which the complete holographic
thermodynamics can be derived, and its only difference with the
familiar photon gas is that we require $\varepsilon=Gp^3$ other than
$\varepsilon=p$. Certainly we cannot expect such a simple theory to
describe all the profound phenomena of quantum gravity. Maybe the
constitutes obeying $\varepsilon=Gp^3$ should only be viewed as the
collective excitations or quasi-particles for the corresponding
systems. In the following, we try to explain the meaning of the
relation $\varepsilon=Gp^3$ from the field-theoretical viewpoint.
Under some assumptions, we shall see that $\varepsilon$ should be
understood as the energy distributed to a DoF. As in the above
calculation for thermodynamics, the gravitational effects are only
reflected by the gravitational constant $G$ and we don't intend to
be involved in a complicated analysis of curved space-time.

\subsection{The explanation of $\varepsilon=Gp^3$}
We begin with a review of the simplest theory for massless scalar
fields. The action is $S = \int {d^4 x\partial ^u \varphi }
\partial _u \varphi$. The Hamiltonian is $ H = \int {d^3 x} [\dot
\varphi ^2 + \left( {\nabla \varphi } \right)^2]$. When confined to
a box of volume $V$, the scalar field can be decomposed as
\begin{align}\varphi = \sum\limits_k {\frac{1}{\sqrt {2w_k V} }\left[ {a_k
e^{i\left( {\vec{k}\cdot\vec{r} - w_k t} \right)}  + a_k ^\dag e^{ -
i\left( {\vec{k}\cdot\vec{r} - w_k t} \right)} }
\right]},\label{phi}
\end{align}
where $w_k=|\vec{k}|$. For a particle explanation, we should require
\begin{align}
\left[ {a_k ,a_l ^\dag  } \right] = \delta _{kl} ,\ \ \  \left[ {a_k
,a_l } \right] = \left[ {a_k ^\dag  ,a_l ^\dag  } \right] =
0.\label{com}
\end{align}
With Eqs.\eqref{phi} and \eqref{com}, the Hamiltonian can be
simplified and expressed as
\begin{align}
H= \sum\limits_k {w_k } a_k ^\dag  a_k .\label{waa}
\end{align}
As known, the theory describes infinite quantum harmonic oscillators
with each particle excitation has energy $w=cp$.

Now we assume a general system can be effectively described by a
field denoted as $\Phi$. Next we require $\Phi$ to be a
dimensionless field, then the action naturally takes the form $S =
\frac{1}{G}\int {d^4 x\partial ^u \Phi } \partial _u \Phi$ by
dimensional analysis. The corresponding Hamiltonian is $H =
\frac{1}{G}\int {d^3 x} [\dot \Phi ^2 + \left( {\nabla \Phi }
\right)^2 ]$. The field can be decomposed as
\begin{align}
\Phi = \sum\limits_k \frac{ G^\alpha {w_k}^{ 2\alpha-\frac{3}{2} }
}{ \sqrt {2V}} \left[ {a_k e^{i\left( {\vec{k}\cdot\vec{r} - w_k t}
\right)}  + a_k ^\dag e^{ - i\left( {\vec{k}\cdot\vec{r} - w_k t}
\right)} } \right].\label{gphi}
\end{align}
The factor $1/ \sqrt V $ is in order to cancel out the integral of
coordinates in the Hamiltonian for a while. The factor $G^\alpha p^{
2\alpha-\frac{3}{2} } $ is a general form insuring $\Phi$ to be a
dimensionless quantity. The value of $\alpha$ is arbitrary here but
it will be discussed later. The commutation relation \eqref{com}
should still hold to insure the physical picture of particle
excitation. Substituting Eqs.\eqref{com} and \eqref{gphi} into the
Hamiltonian, we find
\begin{align}
H= \sum\limits_k G^{2\alpha-1} {w_k }^{4\alpha-1} a_k ^\dag  a_k
.\label{gwaa}
\end{align}
When $\alpha=1/2$ we recover Eq.\eqref{waa}, so the energy
distributed to a DoF is $\varepsilon=w$. It is easy to understand
because $\Phi/\sqrt{G}$ has dimension $1$ and using it as the
fundamental field the theory returns to the standard scalar theory.
In contrast, when $\alpha=1$ we get the expected
\begin{align}
H= \sum\limits_k G {w_k }^{3} a_k ^\dag a_k .
\end{align}
The theory is different from conventional QFT in various aspects.
First, the energy distributed to a DoF is $\varepsilon=G w^{3}$
other than $w$. This gives a field-theoretical explanation of
$\varepsilon=Gp^3$. Note that the energy-momentum relation $w=p$
still holds since $e^{ - i ({\vec{k}\cdot\vec{r} \pm w_k t}) }$ is
the planar wave solution of $\partial ^\mu
\partial _\mu \Phi  = 0$. There are two different energies at hand now. It reminds
that in a gravitational system the energy measured by a local
observer is red-shifted to a distant observer and this effect can
remarkably influence the thermodynamics of a system \cite{pad,us3}.
So it is possible that $w$ could be explained as the intrinsic
energy of the oscillator and $Gw^3$ as the energy measured by an
exterior observer. Second, the theory with $\varepsilon=Gw^3$
possesses the entire holographic DoFs as we analyzed. Obviously the
DoFs increase from $(A/G)^{3/4}$ to $A/G$ as the parameter $\alpha$
changes from $1/2$ to $1$. Third, the standard canonical commutation
relation $ [\Phi(\vec{x}),\pi (\vec{x})] = i$ (together with
$\varepsilon=w$) is only applicable to the conventional QFT case
with $\alpha=1/2$. For the holographic case with $\alpha=1$, a
detailed analysis will lead to $[q,p]=iGp^2$ where the operator $q$
is constructed from $a_k$ and $a_k ^\dag$ and the conjugate $p$ is
extracted from the corresponding Lagrangian. Notably, it has the
same form as the second term of a generalized commutation relation
$[ q,p] = i\left( {1 + Gp^2 } \right)$ which corresponds to the
generalized uncertainty principle (GUP)
\cite{gup1,gup2,gup3,gup4,gup5}. The proposal of GUP $ \Delta
q\Delta p \ge \frac{1}{2}\left( {1 + G\left( {\Delta p} \right)^2 }
\right)$ is mainly to incorporate with the minimum length
$l_p=\sqrt{G}$ of quantum gravity. The momentum $p$ can also be
taken as $p_o+Gp_o^3$ with $[x,p_0]=i$ to realize the GUP
\cite{gup3,gup4}. Again the second term has the same form as our
holographic formula $\varepsilon=Gp^3$, though the exact physical
connection between them is not clear for now.

\subsection{Holographic state space and holographic entropy bound}

In \cite{us1} the dimension of the Hilbert space for the
conventional QFT is found to be $e^{(A/G)^{3/4}}$. The method can be
utilized to analyze the holographic state space.

When confined to a region of size $L$, the particle's momentum
$\vec{p}$ is quantized as $\frac{1}{L} (n_x,n_y,n_z)$. Introducing
an effective ultraviolet momentum cutoff $\Lambda$, the total number
of the quantized modes is $N=\sum\limits_{\vec{k}}1\sim
l^{3}\Lambda^{3}$. Then the state space of the system can be
constructed by acting $a_k^\dag$ in sequence on the vacuum state
$|\Omega\rangle$, that is
\begin{align}
\left( {a_{k_1 } ^\dag  } \right)^{n_1 } \left( {a_{k_2 } ^\dag  }
\right)^{n_2 } \cdots \left( {a_{k_{N} } ^\dag  } \right)^{n_N}
|\Omega \rangle .\label{space}
\end{align}
Different sets of the occupation number $\{n_i\}$ corresponds to
independent quantum states. Now we consider the states satisfying
the gravitational stable requirement $E= n_1 \varepsilon _1  + n_2
\varepsilon _2  + \cdots n_N \varepsilon _N \leq E_{bh}$ as the
physically permitted state, where $\varepsilon=Gp^3$ should be
applied for the holographic case and $E_{bh}=L/G$ is the black hole
energy of the same size. To get the dimension of the physically
permitted Hilbert space, we need to count the total number of these
states.

We start from the simplest states $ \left( {a_{k_{i_1} } ^\dag  }
\right)^{n_{i_1} } \left( {a_{k_{i_2} } ^\dag  } \right)^{n_{i_2} }
|\Omega \rangle$ $(i_1 \ne i_2)$ with only two modes being excited.
The number of states satisfying $ n_{i_1 } \varepsilon _{i_1 }  +
n_{i_2 } \varepsilon _{i_2 }  \leq E_{bh}$ can be evaluated as
$\frac{1}{2!} S_2= \frac{1}{{2!}}\sum\limits_{i_1 < i_2 }^N
{\frac{{E_{bh} }}{{\varepsilon _{i_1 } }}\frac{{E_{bh}
}}{{\varepsilon _{i_2 } }}}$. The calculation is easy to be
generalized to the counting of the states with $m$ modes
simultaneous excited, which is
\begin{align}
\begin{split}
\frac{1}{m!}{S_m } &= \frac{1}{m!}\sum\limits_{i_1  < i_2 \cdots
<i_m }^N {\frac{{E_{bh} }}{{\varepsilon _{i_1 } }}\frac{{E_{bh}
}}{{\varepsilon _{i_2 }
}}\cdots \frac{{E_{bh} }}{\varepsilon _{i_m } }}\\
&<\frac{1}{\left( {m!} \right)^3 }\left( {L^3 \int_0^\Lambda
{\sqrt {\frac{{E_{bh} }}{{Gp^3 }}} } p^2 dp} \right)^{2m}\\
& = \frac{1}{{\left( {m!} \right)^3 }}\left( {\frac{{E_{bh} L^6
\Lambda ^3 }}{G}} \right)^m .
\end{split}
\end{align}
So the dimension of the Hilbert space is
\begin{align}
W= \sum\limits_{m = 1}^N \frac{1}{m!}{S_m }< \sum\limits_{m = 1}^N
{\frac{1}{{\left( {m!} \right)^3 }}} z^m \sim \frac{1}{{2\sqrt 3 \pi
z^{1/3} }}e^{3z^{1/3} }, \label{hilbert}
\end{align}
where $z\equiv\frac{{E_{bh} L^6 \Lambda ^3 }}{G}$. In the above
summation, the state number $\frac{1}{m!} S_m $ peaks at
$m_0=(\frac{{E_{bh} L^6 \Lambda ^3 }}{G})^{1/3}$ and when $m>m_0$
the state number drops dramatically to $0$. On the other hand, there
is surely a physical limitation on the maximum value of excited
modes $m$. The lowest energy state with $m$ modes excited is the
state with one particle occupying one mode successively. The value
of $m$ could be so large that even the state with the lowest energy
has $E>E_{bh}$ and should not contribute to the counting of
physically permitted states. The detailed physical analysis leads to
$m_0=(\frac{ E_{bh} L^3 }{G})^{1/2}$. Then the consistency between
the mathematical and physical scenarios requires $(\frac{{E_{bh} L^6
\Lambda ^3 }}{G})^{1/3}=(\frac{ E_{bh} L^3 }{G})^{1/2}$, which
immediately leads to $\Lambda=(GL)^{-1/3}$ as we expect for a
holographic theory. Substituting $E_{bh}= L/G $ and
$\Lambda=(GL)^{-1/3}$ into Eq.\eqref{hilbert}, we get the dimension
of the Hilbert space $W<e^{A/G}$ and the holographic entropy bound
$S=\ln W<A/G$.

\section{Conclusion}
The conventional QFT is only applicable to the scale $\Lambda\leq
(\sqrt{G} L)^{-1/2}$. In this paper, we suggested a new theory
exists at the scale from $(\sqrt{G} L)^{-1/2}$ to $(GL)^{-1/3}$,
where new DoFs should emerge and the entropy gap from $(A/G)^{3/4}$
to $A/G$ can be overcome. We provided some preliminary thoughts in
this direction. By a thermodynamical analysis we proposed that for
the holographic theory the energy distributed to a microscopic DoF
should be $\varepsilon=Gp^3$. Using this relation rather than
$\varepsilon=p$ as the starting point, the standard statistical
analysis verifies that it leads to the complete behaviors of
holographic thermodynamics: $E \sim L/G$, $T \sim 1/L$ and $S \sim
A/G$. It furthermore gives the equation of state $\rho=p$ for the
holographic constitutes, which happens to be consistent with the
cosmological holographic entropy bound. Finally, we have tried to
give a field-theoretical explanation of $\varepsilon=Gp^3$ and
discussed several differences between the theory and the
conventional QFT. Using the field theory viewpoint, we also
constructed the state space of the holographic theory and derived
the holographic bound.

Though these thoughts are quite rough, it may still shed some light
on the understanding of holographic principle and quantum properties
of gravity. Here we want to stress that it can also provide new
ideas to the understanding of the early stage of the universe. When
we gradually trace back to the early stage of the universe, we
encounter higher and higher energy scale physics from atomic physics
to nuclear physics and to grand unified physics. In this spirit, our
work strongly suggests a holographic stage of the universe before
the conventional quantum fields dominated stage. Even for the
earlier inflation stage, the holographic eternal inflation model
proposed in \cite{bf4} serves as a good choice. So we expect the
universe starts from a holographic stage. Afterwards the density of
the holographic fluid with $w=\rho/p=1$ will be diluted by
conventional constitutes with $w=1/3$ (radiation) and $w=0$
(matter), since the cosmological evolution favors to lower the value
of $w$. We hope some kinds of remnant indications of this
holographic stage could be detected in future experiment.

\section*{Acknowledgements} I would like to thank Y.S. Song for
useful discussions. The work was first oral reported at the
Institute of Theoretical Physics of Lanzhou University when I
visited there in August, 2017. I gratefully acknowledge the
hospitality and the useful suggestions I received there. The work is
supported in part by the NNSF of China with Grant No. 11205045, the
NSF of Hebei province with Grant No. A2015201212 and the NSF of
Hebei Eduction Department with Grant No. YQ2014032.


\begin{thebibliography}{99}
\bibitem{note} We prove that $\rho=p$ can be derived from a holographic thermodynamcs.
 But the converse statement that $\rho=p$ leads to a holographic theroy is not always correct.
To fit with $TdS=dE+pdv\sim L/G$, both the case $T=(\sqrt{G}
L)^{-1/2}$, $S=(A/G)^{3/4}$ and the case $T=1/L$, $S=A/G$ can be
applied nicely, depending on the concrete microscopic picture of the
theory. Thus, as a conventional QFT, the scalar field theory with
$\rho=p$ should still obey the $S=(A/G)^{3/4}$ entropy bound.
\end{thebibliography}
\end{document}